\documentclass[12pt,preprint]{emulateapj}
\usepackage{graphicx}

\shorttitle{The 70 Vir Planetary System}
\shortauthors{Stephen R. Kane et al.}
\slugcomment{Submitted for publication in the Astrophysical Journal}

\begin{document}

\title{A Comprehensive Characterization of the 70 Virginis Planetary
  System}

\author{
  Stephen R. Kane\altaffilmark{1},
  Tabetha S. Boyajian\altaffilmark{2},
  Gregory W. Henry\altaffilmark{3},
  Y. Katherina Feng\altaffilmark{4,5,6},
  Natalie R. Hinkel\altaffilmark{1},
  Debra A. Fischer\altaffilmark{2},
  Kaspar von Braun\altaffilmark{7},
  Andrew W. Howard\altaffilmark{8},
  Jason T. Wright\altaffilmark{4,5}
}
\email{skane@sfsu.edu}
\altaffiltext{1}{Department of Physics \& Astronomy, San Francisco
  State University, 1600 Holloway Avenue, San Francisco, CA 94132,
  USA}
\altaffiltext{2}{Department of Astronomy, Yale University, New Haven,
  CT 06511, USA}
\altaffiltext{3}{Center of Excellence in Information Systems, Tennessee
  State University, 3500 John A. Merritt Blvd., Box 9501, Nashville,
  TN 37209, USA}
\altaffiltext{4}{Department of Astronomy and Astrophysics,
  Pennsylvania State University, 525 Davey Laboratory, University
  Park, PA 16802, USA}
\altaffiltext{5}{Center for Exoplanets \& Habitable Worlds,
  Pennsylvania State University, 525 Davey Laboratory, University
  Park, PA 16802, USA}
\altaffiltext{6}{Department of Astronomy \& Astrophysics, 1156 High
  Street, MS: UCO / LICK, University of California, Santa Cruz, CA
  95064, USA}
\altaffiltext{7}{Lowell Observatory, 1400 West Mars Hill Road,
  Flagstaff, Arizona 86001, USA}
\altaffiltext{8}{Institute for Astronomy, University of Hawaii,
  Honolulu, HI 96822, USA}


\begin{abstract}

An on-going effort in the characterization of exoplanetary systems is
the accurate determination of host star properties. This effort
extends to the relatively bright host stars of planets discovered with
the radial velocity method. The Transit Ephemeris Refinement and
Monitoring Survey (TERMS) is aiding in these efforts as part of its
observational campaign for exoplanet host stars. One of the first
known systems is that of 70 Virginis, which harbors a jovian planet in
an eccentric orbit. Here we present a complete characterization of
this system with a compilation of TERMS photometry, spectroscopy, and
interferometry. We provide fundamental properties of the host star
through direct interferometric measurements of the radius (1.5\%
uncertainty) and through spectroscopic analysis. We combined 59 new
Keck HIRES radial velocity measurements with the 169 previously
published from the ELODIE, Hamilton, and HIRES spectrographs, to
calculate a refined orbital solution and construct a transit ephemeris
for the planet. These newly determined system characteristics are used
to describe the Habitable Zone of the system with a discussion of
possible additional planets and related stability
simulations. Finally, we present 19 years of precision robotic
photometry that constrain stellar activity and rule out central
planetary transits for a Jupiter-radius planet at the 5$\sigma$ level,
with reduced significance down to an impact parameter of $b = 0.95$.

\end{abstract}



\section{Introduction}
\label{introduction}

The exoplanet discoveries over the past couple of decades have
revealed a particular need to understand the properties of the host
stars. This is because the planetary parameters derived from the
detection methods of radial velocities (RV) and transits rely heavily
upon the mass and radius determinations of their parent stars. These
are often determined using stellar models, but there are ongoing
efforts to provide more direct measurements of the stellar properties
through asteroseismology \citep{hub14} and interferometry
\citep{2012ApJ...757..112B,2013ApJ...771...40B,von14}. The importance
of these measurements cannot be overstated since they not only affect
the derived planetary properties but also the quantification of the
Habitable Zone (HZ) \citep{kas93,kop13b,kop14} and subsequent
calculations of the fraction of stars with Earth-size planets in the
HZ, or $\eta_{\oplus}$ \citep{dre13,kop13a,pet13}.

The Transit Ephemeris Refinement and Monitoring Survey (TERMS) is
aiding in stellar characterization for the brightest stars as part of
its program to improve exoplanetary orbital parameters
\citep{kan08,kan09}. Recent results include detailed spectroscopic and
photometric analyses of the planet-hosting stars HD~38529
\citep{hen13} and HD~192263 \citep{dra12} and the identification of
long-term activity cycles. TERMS observations have also led to the
discovery of new planets in the HD~37605 \citep{wan12} and HD~4203
\citep{kan14b} systems. These efforts are continuing with a focus on
the brightest host stars, which tend to be those around which planets
were discovered using the RV method.

One of the earliest exoplanet discoveries was that of the planet
orbiting the bright ($V = 5$) star 70 Virginis (hereafter 70~Vir). The
planet was discovered by \citet{mar96} and lies in an eccentric ($e =
0.4$) 116~day orbit. \citet{per96} subsequently used {\it Hipparcos}
astrometry to constrain the inclination and thus determine that the
companion is indeed sub-stellar in mass. Observations of 70~Vir have
continued since discovery, with RV data from Observatoire de
Haute-Provence published by \citet{nae04} and the complete Lick
Observatory dataset compiled by \citet{but06}. A detailed
characterization of the nearest and brightest exoplanet host stars is
important because these continue to be those which are most suitable
for potentially studying exoplanetary atmospheres for transiting
planets.

\begin{deluxetable*}{rccc}
\tablecaption{Log of interferometric observations for
  70~Vir \label{tab:interferometric_observations}}
\tablewidth{0pc}
\tablehead{
\colhead{UT Date}     &
\colhead{Baseline} &
\colhead{\# of Obs} &
\colhead{Calibrators}
}
\startdata
2013/04/03 & S1/E1 & 5 & HD\,119288, HD\,121560, HD\,122386 \\
2013/04/04 & S1/E1 & 4 & HD\,119550, HD\,113022 \\ 
2014/04/20 & W1/E1 & 4 & HD\,119288, HD\,119550
\enddata
\tablecomments{Calibrator angular diameters from
  JSDC\footnote{Available at http://www.jmmc.fr/searchcal} are:
  $\theta_{\rm HD113022}=0.39 \pm 0.03$, $\theta_{\rm HD119288}=0.39
  \pm 0.03$, $\theta_{\rm HD119550}=0.38 \pm 0.03$, $\theta_{\rm
    HD121560}=0.46 \pm 0.03$, and $\theta_{\rm HD122386}=0.49 \pm
  0.03$~mas. For details on the interferometric observations, see
  \S\ref{interobs}.}
\end{deluxetable*}

Here we provide a detailed analysis of the 70~Vir system for both the
star and the known planet. Section \ref{interobs} describes new
interferometric observations obtained using the CHARA Array. Section
\ref{stellar} combines these measurements with a new spectroscopic
analysis of Keck/HIRES data to determine fundamental stellar
properties of 70~Vir. Section \ref{orbit} presents the addition of
$\sim 60$ Keck/HIRES RV measurements to the existing time series, a
revised Keplerian orbital solution, and the calculation of an accurate
transit ephemeris. Section \ref{hz} uses the greatly improved system
parameters to calculate the extent of the HZ and discusses the
prospect of HZ planets in the system. Section \ref{photometry}
describes 19 years of precision robotic photometry that both rule out
a planetary transit and show that the long-term stellar activity is
constant within 0.004 mag. We provide concluding remarks in Section
\ref{conclusions}.


\section{Interferometric observations}
\label{interobs}

70~Vir (HD~117176; HR~5072; HIP~65721) is a bright \citep[$V=4.97$;
  $H=3.24$;][]{1968ApJ...152..465J} and nearby \citep[{\it Hipparcos}
  parallax of $55.60 \pm 0.24$~mas;][]{2007A&A...474..653V} star. Our
interferometric observations of 70~Vir were conducted at the Georgia
State University's Center for High Angular Resolution Astronomy
(CHARA) Array and the Classic beam combiner in two-telescope mode
operating in $H$-band (central wavelength $\lambda_{\rm c} = 1.67
\mu$m) \citep{2005ApJ...628..453T}. We collected a total of 13
observations over the course of three nights: two nights in April 2013
using the S1E1 pair of telescopes, and one night in April 2014 using
the W1E1 pair of telescopes.  The S1E1 and E1W1 are the two longest
telescope configurations available at CHARA, with baselines $B$
(distances between two telescopes) of $B_{\rm S1E1} = 330$~m and
$B_{\rm W1E1} = 313$~m.

Calibrator stars were observed in bracketed sequences with 70~Vir. We
use the SearchCal software
\citep{2006A&A...456..789B,2011A&A...535A..53B} to select calibrator
stars based on their proximity in the sky with respect to the science
target (within $\sim 8$~degrees). We observe a total of five
calibrators with estimated angular sizes $\theta_{\rm est} < 0.5$~mas
in order to minimize systematic errors which could be introduced by
the calibrator's estimated sizes \citep{2005PASP..117.1263V}.  A log
of the observations along with calibrator information can be found in
Table~\ref{tab:interferometric_observations}.

\begin{figure}
  \includegraphics[width=8.2cm]{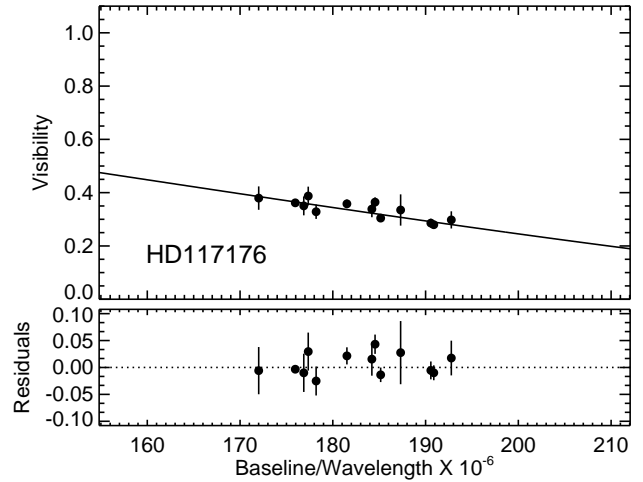} 
  \caption{Plot of calibrated interferometric visibilities and the
    limb-darkened angular diameter fit for 70~Vir. The interferometric
    observations are described in
    Section~\ref{interobs}.}
  \label{fig:visibilities}
\end{figure}

Calibrated data are used to determine the stellar uniform disk angular
diameter $\theta_{\rm UD}$ and limb darkened angular diameter
$\theta_{\rm LD}$ by fitting the functions expressed in
\citet{1974MNRAS.167..121H}. An estimate of the star's temperature and
gravity based on spectra are used to determine $H$-band limb-darkening
coefficients from \citet{2011A&A...529A..75C}.  These quantities are
iterated upon with the final stellar parameters (see Section
\ref{stellar}) to determine the final coefficient used in the limb
darkened diameter solution, $\mu_{\rm H} = 0.3512$ (e.g., see
\citealt{2012ApJ...757..112B,2013ApJ...771...40B}). We measure the
angular diameter of 70~Vir to be $\theta_{\rm UD} = 0.967\pm0.004$ and
$\theta_{\rm LD} = 0.998 \pm
0.005$~milliarcseconds. Figure~\ref{fig:visibilities} shows the
interferometric data along with the best fit limb-darkened visibility
function; the angular diameter measurements may be found in
Table~\ref{tab:stellar_properties}.

\begin{deluxetable*}{rccc}
\tablecaption{Stellar properties for 70~Vir \label{tab:stellar_properties}} 
\tablewidth{0pc}
\tablehead{
\colhead{ } &
\colhead{Value} &
\colhead{Value}	&
\colhead{ }	\\
\colhead{Parameter}	&
\colhead{Spectroscopic}	&
\colhead{Interferometric} &
\colhead{Section Reference}		
}
\startdata
$\theta_{\rm UD}$ (mas) \dotfill		&	\nodata	&	$0.967\pm0.004$	&	\S \ref{interobs}	\\
$\theta_{\rm LD}$ (mas)	\dotfill	&		\nodata	&	$0.998\pm0.005$	&	\S \ref{interobs}	\\
$F_{\rm bol}$ ($10^{-8}$~erg~s$^{-1}$~cm$^{-2}$)	&	\nodata	&	$28.050	\pm 0.562$\phn	& \S \ref{sec:radius_teff}	\\
Luminosity (L$_{\rm \odot}$) \dotfill	& \nodata 	&	$2.827 \pm 0.062$	&	\S \ref{sec:radius_teff}	\\
Radius $R_*$ (R$_{\rm \odot}$) \dotfill	&	$1.94 \pm 0.05$	&	$1.9425 \pm 0.0272$ 	&	\S \ref{sec:sme}, \S \ref{sec:radius_teff}\\
$T_{\rm eff}$ (K)	\dotfill			&	$5439 \pm 44$\phn\phn	&	$5393 \pm 30$\phn\phn		& \S \ref{sec:radius_teff} \\ 
$[$Fe/H$]$ \dotfill		&	$-0.09 \pm 0.03$\phs	&	\nodata	&	\S \ref{sec:sme}	\\
$v \sin i$ (km~s$^{-1}$)	\dotfill	&	$1.56 \pm 0.50$	&	\nodata	&	\S \ref{sec:sme}	\\
$\log g$ \dotfill	&	$3.90 \pm 0.06$	&	\nodata 	&	 \S \ref{sec:sme}	\\
Mass $M_*$ (M$_{\rm \odot}$) \dotfill	&	$1.09 \pm 0.02$		&	\nodata 	&	\S \ref{sec:sme}	\\
Age (Gyr)	\dotfill	& $7.77 \pm 0.51$	&	\nodata	&	\S \ref{sec:sme}
\enddata
\tablecomments{For details, see \S \ref{stellar}.}
\end{deluxetable*}


\section{Stellar Properties}
\label{stellar}


\subsection{Interferometry}
\label{sec:radius_teff}

The stellar angular diameter measured with interferometry in
Section~\ref{interobs} may be used in combination with the
trigonometric parallax from {\it Hipparcos} to derive the physical
linear radius of the star, $R$, using trigonometry. The most direct
way to measure the effective temperature of a star is via the
Stefan-Boltzmann equation, $L = 4 \pi R^2 \sigma T_{\rm eff}^4$,
rearranged to yield
\begin{equation}\label{eq:teff}
T_{\rm eff} = 2341 (F_{\rm bol} / \theta_{\rm LD}^2)^{0.25},
\end{equation}
where the constant 2341 absorbs the conversion constants
assuming using units of $F_{\rm bol}$ in
$10^{-8}$~erg~s$^{-1}$~cm$^{-2}$ and the limb-darkened angular
diameter $\theta_{\rm LD}$ in milliarcseconds (mas).

The bolometric flux is measured by normalizing a G5 V spectral
template from the \citet{1998PASP..110..863P} library to broad-band
photometry and spectrophotometry in the literature. Details of this
method are described in \citet{2008ApJS..176..276V,von14}. For 70~Vir,
we use photometry from the following references:
\citet{1953ApJ...117..313J, 1966PDAUC...1....1G, 1966CoLPL...4...99J,
  1967AJ.....72.1334C, 1966ZA.....64..116P, 1966PASP...78..546J,
  1976HelR....1....0P, 1954ApJ...120..196J, 1963MNRAS.125..557A,
  1961LowOB...5..157S, 1966ArA.....4..137H, 1985A&AS...61..331O,
  1986EgUBV........0M, 1975MNRAS.172..667J, 1979PASP...91..180M,
  2002yCat.2237....0D, 1968ApJ...152..465J, 1988iras....1.....B,
  2003tmc..book.....C, 1999yCat.2225....0G, 1996BaltA...5..103O,
  1981SAAOC...6...10D, 1994A&AS..106..257O, 1990VilOB..85...50J,
  1988A&A...206..357R, 1970A&AS....1..199H, 1991TrSht..63....4K,
  1975RMxAA...1..299J, 2004ApJS..154..673S}.  We also use the
spectrophotometry data from \citet{1988scsb.book.....K,
  1998yCat.3207....0G, 1985BCrAO..66..152B}.

\begin{figure}
  \includegraphics[width=8.2cm]{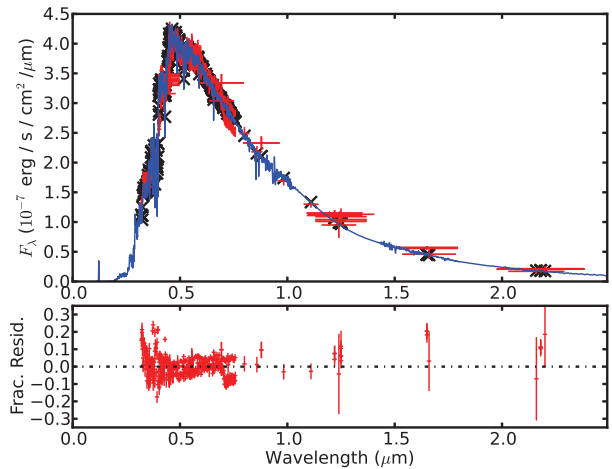} 
  \caption{{\it Upper panel}: (1) the blue curve is a G5 V spectral
    template from the \citet{1998PASP..110..863P} library, (2) the red
    crosses are literature photometry and spectrophotometry data of
    70~Vir with errors in y-direction and filter bandwidths in
    x-direction, (3) the black X-shapes represent the specific flux
    values of the spectral template at the central wavelength of the
    filter of the respective literature photometry data point. {\it
      Lower panel}: the red crosses represent the fractional residuals
    to the fit. The high density of photometry points toward the blue
    end of the spectrum stems from the three spectrophotometry data
    sets in the literature for this star. For more details, see
    Section~\ref{sec:radius_teff}.}
  \label{fig:sed}
\end{figure}

The fit (see Figure \ref{fig:sed}) produces a bolometric flux $F_{\rm
  bol} = (28.050 \pm 0.0248) \times 10^{-8}$~erg~s$^{-1}$~cm$^{-2}$.
We note that the quoted uncertainty is statistical only and thus does
not account for absolute errors in the templates, uncertain
photometric zero-points, or other effects such as the ones outlined in
Section 2.2 of \citet{von14}. We follow the arguments in Sections
3.2.1 and 3.2.2 of \citet{boh14} and add a 2\% error in quadrature to
account for a more realistic representation of the true uncertainties
\cite[see also the appendix in][]{2012PASP..124..140B}. The final
$F_{\rm bol}$ and associated uncertainty values are presented in
Table~\ref{tab:stellar_properties}, along with the $T_{\rm eff}$
derived from Equation~\ref{eq:teff} using $F_{\rm bol}$ and
$\theta_{\rm LD}$.

\subsection{Spectroscopy}
\label{sec:sme}

We model two spectra of 70~Vir, taken on 3 July 2009 with Keck/HIRES.
The spectra are modeled using the Spectroscopy Made Easy (SME) package
\citep{val96,2005ApJS..159..141V}. SME employs an iterative mode using
results of the model atmosphere analysis in combination with the
Yonsei-Yale model isochrones \citep{2004ApJS..155..667D} in order to
produce self-consistent results with the measured surface gravity
\citep{2009ApJ...702..989V}. The results of the spectroscopic modeling
(surface gravity $\log g$, rotational velocity $v \sin i$, atmospheric
abundance [Fe/H], and effective temperature $T_{\rm eff}$) and stellar
isochrone solution (mass $M_*$, radius $R_*$, and age) are presented
in Table~\ref{tab:stellar_properties}. The values for effective
temperature measured with interferometry and derived with
spectroscopic modeling agree well. Likewise, there is excellent
agreement with the radius predicted by model isochrones and the
directly measured radius with interferometry. The introduction of the
interferometric data set not only allows for an empirically based
consistency check with the results from stellar atmosphere and
evolutionary codes, but also reduces the uncertainties in the stellar
parameters beyond the capabilities of current methods employing models
($\sigma T_{\rm eff}$ and $\sigma R_{\ast}$ are 85\% and 47\% lower
respectively).


\subsection{Stellar Abundances}
\label{sec:abund}

There are at least 17 different groups who have measured the stellar
abundances in 70~Vir, for example \citet{zha02} and
\citet{das11}. Given the close proximity of the host star ($\sim$18pc)
and bright $V$ magnitude, elements from lithium to europium have been
measured within 70~Vir. Per the analysis of \citet{hin14}, the
abundance measurements as determined by each group were renormalized
to the \citet{lod09} solar abundance scale. The maximum variation
between datasets for each element, or the {\it spread}, was determined
in order to better characterize the consistent measurement of that
element abundance. For 70~Vir, the median value for [Fe/H] = -0.01
dex, while the spread was 0.41 dex, since the renormalized
\citet{lai85} determination found [Fe/H] = -0.2 dex and renormalized
\citet{val08} determined [Fe/H] = 0.2 dex. In the analysis of the {\it
  Hypatia Catalog} \citep{hin14}, to ensure abundances were copacetic,
a star was not considered when the spread of the catalog abundances
was greater than the error bar associated with that element, in this
case $\pm$ 0.05 dex.  Therefore, because of the discrepancy in [Fe/H]
between datasets, 70 Vir was not incorporated into the analysis of
{\it Hypatia}.

Despite the $\sim$30 elements that have been measured in 70~Vir, the
majority of them suffer from inconsistent measurements between groups.
The only elements that do not have a spread greater than the
respective error bar are for cases where only one literature source
has measured that element, or when the spread is 0.0 dex.  In these
cases, the renormalized abundances are [Li/Fe] = -1.54 dex and [Eu/Fe]
= 0.01 dex \citep{gon07}; [K/Fe] = -0.21 dex \citep{zha02}; [ZrII/Fe]
= -0.13 \citep{mas00}; [VII/Fe] = -0.09 dex and [CrII/Fe] = -0.42 dex
\citep{tak07}; and [Sr/Fe] = 0.04 dex, [Zr/Fe] = 0.09 dex, [CeII/Fe] =
0.08 dex \citep{das11}. While many of these abundances are sub-solar,
not much can be said given their varying nucleosynthetic origins.


\section{A Refined Planetary Orbit}
\label{orbit}

Here we present new RV data for 70~Vir, a revised Keplerian orbital
solution for the planet, and an accurate transit ephemeris for 70~Vir.


\subsection{Spectra Acquisition}
\label{sec:spectra}

Previously published data for 70~Vir includes 169 measurements acquired
with the Hamilton Echelle Spectrograph \citep{vog87} on the 3.0m Shane
Telescope at Lick Observatory \citep{mar96,but06,fis14} and 35
measurements acquired with the ELODIE spectrograph \citep{bar96} on
the 1.93m telescope at Observatoire de Haute-Provence
\citep{nae04}. We add to these time series 59 new measurements
acquired with the HIRES echelle spectrometer \citep{vog94} on the
10.0m Keck I telescope. H\&K emission measured from the Keck spectra
show that 70~Vir is a relatively quiet star. We show the complete
dataset of 263 measurements in Table \ref{rvs}, where the fourth
column indicates the source of the measurements. These data represent
a baseline of $\sim 26$~years of monitoring 70~Vir.

\LongTables
\begin{deluxetable}{cccc}
  \tablewidth{0pc}
  \tablecaption{\label{rvs} 70~Vir Radial Velocities}
  \tablehead{
    \colhead{Date} &
    \colhead{RV} &
    \colhead{$\sigma$} &
    \colhead{Telescope} \\
    \colhead{(JD -- 2,440,000)} &
    \colhead{(m\,s$^{-1}$)} &
    \colhead{(m\,s$^{-1}$)} &
    \colhead{Instrument}
  }
  \startdata
7195.02270 & -155 & 13 & Hamilton \\
7195.02790 & -146 & 13 & Hamilton \\
7195.03310 & -146 & 13 & Hamilton \\
7224.84280 & 145 & 10 & Hamilton \\
7224.84810 & 146.0 & 9.6 & Hamilton \\
7224.85300 & 144 & 10 & Hamilton \\
7373.71740 & 109.7 & 9.4 & Hamilton \\
7373.72130 & 115.1 & 9.5 & Hamilton \\
7373.72640 & 114.7 & 8.5 & Hamilton \\
7578.89680 & 232.3 & 9.3 & Hamilton \\
7710.75280 & 429 & 11 & Hamilton \\
7965.00190 & 4 & 12 & Hamilton \\
8018.84930 & -130 & 13 & Hamilton \\
8376.80300 & -66 & 12 & Hamilton \\
8437.74980 & -55 & 12 & Hamilton \\
8670.98610 & -73 & 11 & Hamilton \\
8670.99340 & -82 & 14 & Hamilton \\
8746.68150 & 254 & 14 & Hamilton \\
8746.70340 & 291 & 13 & Hamilton \\
8781.77040 & -12 & 11 & Hamilton \\
8781.79240 & 4 & 10 & Hamilton \\
8847.67430 & -37 & 11 & Hamilton \\
8993.08360 & 442.3 & 9.9 & Hamilton \\
9068.91280 & -110 & 10 & Hamilton \\
9068.93410 & -110.3 & 9.8 & Hamilton \\
9096.85900 & 263 & 10 & Hamilton \\
9096.88040 & 274 & 11 & Hamilton \\
9124.77090 & 94 & 10 & Hamilton \\
9124.79250 & 98 & 11 & Hamilton \\
9172.70420 & -141 & 11 & Hamilton \\
9349.09110 & 305 & 11 & Hamilton \\
9411.97090 & -128 & 11 & Hamilton \\
9469.81290 & 216 & 13 & Hamilton \\
9767.00320 & -120.8 & 5.2 & Hamilton \\
9768.97949 & -119.3 & 5.6 & Hamilton \\
9793.94141 & 176.7 & 4.7 & Hamilton \\
9793.95117 & 187.2 & 4.2 & Hamilton \\
9793.96191 & 185.4 & 4.2 & Hamilton \\
9793.97168 & 188.2 & 3.9 & Hamilton \\
10087.93164 & -162.9 & 4.4 & Hamilton \\
10089.01953 & -162.8 & 4.5 & Hamilton \\
10089.97363 & -157.4 & 3.7 & Hamilton \\
10091.08594 & -163.3 & 4.4 & Hamilton \\
10121.02539 & -113.9 & 5.5 & Hamilton \\
10121.04785 & -105.5 & 6.1 & Hamilton \\
10122.01562 & -108.8 & 7.3 & Hamilton \\
10122.03711 & -115.9 & 6.3 & Hamilton \\
10124.98535 & -83.0 & 4.6 & Hamilton \\
10125.00977 & -92.9 & 4.4 & Hamilton \\
10125.86133 & -84.6 & 5.4 & Hamilton \\
10127.01758 & -77.8 & 6.5 & Hamilton \\
10127.06055 & -78.8 & 6.9 & Hamilton \\
10129.03223 & -52.2 & 4.6 & Hamilton \\
10129.05469 & -56.4 & 4.8 & Hamilton \\
10144.95215 & 206.6 & 3.8 & Hamilton \\
10145.94727 & 231.7 & 5.8 & Hamilton \\
10145.95996 & 242 & 10 & Hamilton \\
10148.95410 & 320.3 & 5.6 & Hamilton \\
10148.98731 & 316.2 & 5.8 & Hamilton \\
10150.92188 & 386.5 & 4.8 & Hamilton \\
10150.94434 & 367.0 & 4.7 & Hamilton \\
10172.90918 & 133.3 & 4.6 & Hamilton \\
10172.92969 & 139.5 & 8.7 & Hamilton \\
10173.88769 & 111.6 & 5.1 & Hamilton \\
10179.78027 & 23.8 & 4.1 & Hamilton \\
10180.76367 & 7.5 & 3.8 & Hamilton \\
10180.76953 & 4.8 & 4.3 & Hamilton \\
10186.82031 & -41.7 & 5.3 & Hamilton \\
10186.84375 & -52.4 & 5.9 & Hamilton \\
10199.80957 & -132.2 & 5.0 & Hamilton \\
10199.83203 & -132.4 & 4.9 & Hamilton \\
10200.82910 & -142.9 & 4.8 & Hamilton \\
10200.85059 & -143.7 & 5.4 & Hamilton \\
10201.81641 & -135.7 & 5.5 & Hamilton \\
10201.83887 & -142.2 & 5.9 & Hamilton \\
10202.81836 & -144.4 & 5.3 & Hamilton \\
10202.84082 & -152.1 & 4.7 & Hamilton \\
10214.76856 & -172.6 & 5.0 & Hamilton \\
10214.79102 & -171.4 & 5.1 & Hamilton \\
10234.79102 & -128.4 & 5.7 & Hamilton \\
10234.81250 & -118.0 & 5.5 & Hamilton \\
10263.66894 & 261.4 & 3.4 & Hamilton \\
10502.97852 & 412.0 & 4.7 & Hamilton \\
10537.88281 & -72.0 & 4.4 & Hamilton \\
10563.75195 & -154.4 & 4.8 & Hamilton \\
10614.75391 & 288.4 & 5.2 & Hamilton \\
10614.75879 & 283.9 & 4.7 & Hamilton \\
10655.68848 & -80.3 & 4.9 & Hamilton \\
10794.06055 & -162.7 & 3.8 & Hamilton \\
10978.75684 & 395.0 & 5.8 & Hamilton \\
10978.77148 & 408.4 & 5.2 & Hamilton \\
11005.70801 & -81.5 & 5.8 & Hamilton \\
11207.01269 & 466.0 & 5.2 & Hamilton \\
11303.82129 & 43.7 & 5.6 & Hamilton \\
11700.72559 & -41.3 & 9.5 & Hamilton \\
11700.74805 & -31.2 & 6.8 & Hamilton \\
11946.02832 & -124.0 & 4.7 & Hamilton \\
11969.94727 & -157.3 & 5.8 & Hamilton \\
12041.82715 & 94.9 & 6.0 & Hamilton \\
12041.83301 & 90.6 & 5.4 & Hamilton \\
12055.77441 & -82.7 & 5.4 & Hamilton \\
12055.80371 & -76.9 & 5.0 & Hamilton \\
12072.70117 & -164.1 & 5.9 & Hamilton \\
12072.73047 & -153.7 & 6.0 & Hamilton \\
12122.69629 & 71.7 & 4.8 & Hamilton \\
12333.97461 & -122.3 & 6.0 & Hamilton \\
12334.95410 & -121.8 & 6.9 & Hamilton \\
12335.96387 & -123.3 & 4.8 & Hamilton \\
12338.95898 & -99.6 & 5.7 & Hamilton \\
12345.93164 & -55.4 & 5.7 & Hamilton \\
12345.95606 & -53.0 & 6.3 & Hamilton \\
12348.95215 & -25.0 & 6.4 & Hamilton \\
12348.97656 & -29.9 & 4.9 & Hamilton \\
12427.76367 & -176.5 & 5.2 & Hamilton \\
12428.80859 & -178.1 & 4.3 & Hamilton \\
12449.72754 & -141.5 & 6.1 & Hamilton \\
12449.73047 & -134.8 & 5.8 & Hamilton \\
12723.92578 & 465.2 & 5.6 & Hamilton \\
12723.94922 & 459.0 & 6.1 & Hamilton \\
12796.69531 & -154.1 & 4.3 & Hamilton \\
12796.71777 & -152.5 & 4.5 & Hamilton \\
12796.74023 & -147.3 & 4.0 & Hamilton \\
13021.07031 & -166.8 & 5.2 & Hamilton \\
13022.02494 & -166.9 & 5.0 & Hamilton \\
13069.87695 & 400.5 & 4.3 & Hamilton \\
13080.93945 & 367.0 & 4.7 & Hamilton \\
13080.96191 & 360.2 & 4.2 & Hamilton \\
13102.86621 & -51.7 & 4.8 & Hamilton \\
13130.82324 & -171.7 & 4.3 & Hamilton \\
13130.82617 & -166.9 & 5.1 & Hamilton \\
13162.69434 & -72.8 & 4.5 & Hamilton \\
13162.71875 & -61.4 & 4.9 & Hamilton \\
13363.05150 & -165.9 & 4.7 & Hamilton \\
13389.06953 & -111.7 & 4.7 & Hamilton \\
13391.00264 & -106.2 & 5.0 & Hamilton \\
13392.04633 & -97.8 & 4.6 & Hamilton \\
13393.02009 & -92.3 & 4.7 & Hamilton \\
13403.04659 & 8.6 & 4.6 & Hamilton \\
13436.90184 & 205.2 & 5.2 & Hamilton \\
13438.93611 & 150.6 & 5.1 & Hamilton \\
13440.88377 & 122.2 & 5.5 & Hamilton \\
13441.98706 & 100.5 & 4.8 & Hamilton \\
13475.82543 & -145.5 & 4.8 & Hamilton \\
13476.79240 & -152.8 & 5.1 & Hamilton \\
13477.81099 & -161.7 & 4.9 & Hamilton \\
13478.84384 & -162.9 & 4.9 & Hamilton \\
13479.85233 & -159.4 & 4.9 & Hamilton \\
13501.82814 & -138.0 & 5.1 & Hamilton \\
13565.71260 & -17.6 & 5.5 & Hamilton \\
13753.02186 & 24.3 & 7.3 & Hamilton \\
13756.07980 & 66.1 & 5.2 & Hamilton \\
13773.99421 & 448.6 & 4.9 & Hamilton \\
13843.87891 & -148.6 & 4.4 & Hamilton \\
14134.03032 & 276.3 & 4.5 & Hamilton \\
14169.93907 & -150.6 & 4.8 & Hamilton \\
14196.85956 & -149.1 & 4.6 & Hamilton \\
14219.84280 & 24.7 & 4.9 & Hamilton \\
14253.73413 & 199.8 & 5.2 & Hamilton \\
14253.73740 & 191.8 & 4.9 & Hamilton \\
14253.74064 & 198.9 & 5.0 & Hamilton \\
14254.73845 & 183.6 & 4.9 & Hamilton \\
14547.96363 & -155.9 & 4.9 & Hamilton \\
14574.87404 & 98.5 & 4.8 & Hamilton \\
14864.06816 & -135.3 & 6.9 & Hamilton \\
14864.07598 & -131.4 & 7.6 & Hamilton \\
15229.03541 & -165.0 & 7.2 & Hamilton \\
15229.03803 & -165.1 & 7.3 & Hamilton \\
15312.79659 & 46.8 & 4.4 & Hamilton \\
15312.79906 & 47.0 & 4.2 & Hamilton \\
10150.56740 & 399.0 & 7.0 & ELODIE \\
10207.47330 & -124.0 & 7.0 & ELODIE \\
10212.49560 & -121.0 & 7.0 & ELODIE \\
10263.35220 & 295.0 & 7.0 & ELODIE \\
10267.36290 & 411.0 & 8.0 & ELODIE \\
10477.67260 & -19.0 & 7.0 & ELODIE \\
10532.52620 & 28.0 & 7.0 & ELODIE \\
10535.54120 & -3.0 & 7.0 & ELODIE \\
10557.54150 & -104.0 & 7.0 & ELODIE \\
10561.51440 & -129.0 & 7.0 & ELODIE \\
10582.41520 & -90.0 & 7.0 & ELODIE \\
10587.41230 & -62.0 & 7.0 & ELODIE \\
10623.40340 & 506.0 & 7.0 & ELODIE \\
10858.59380 & 515.0 & 7.0 & ELODIE \\
10885.57180 & 0.0 & 8.0 & ELODIE \\
10940.42360 & -52.0 & 7.0 & ELODIE \\
10967.38640 & 395.0 & 7.0 & ELODIE \\
11024.34770 & -121.0 & 8.0 & ELODIE \\
11239.62410 & -32.0 & 7.0 & ELODIE \\
11300.41400 & 38.0 & 7.0 & ELODIE \\
11325.41920 & 505.0 & 7.0 & ELODIE \\
11588.57810 & -20.0 & 7.0 & ELODIE \\                        
11623.51300 & -103.0 & 7.0 & ELODIE \\
11653.51820 & 88.0 & 7.0 & ELODIE \\
11692.42330 & 135.0 & 7.0 & ELODIE \\
11956.64950 & -94.0 & 7.0 & ELODIE \\
12038.51990 & 208.0 & 7.0 & ELODIE \\
12356.57740 & 133.0 & 7.0 & ELODIE \\
12361.58960 & 246.0 & 7.0 & ELODIE \\
12413.50440 & -73.0 & 8.0 & ELODIE \\
12719.55560 & 446.0 & 7.0 & ELODIE \\
12748.52720 & 50.0 & 7.0 & ELODIE \\
12751.53870 & 10.0 & 7.0 & ELODIE \\
12772.41720 & -103.0 & 7.0 & ELODIE \\
12776.49050 & -105.0 & 7.0 & ELODIE \\
13933.82196 & -217.7 & 1.0 & HIRES \\
13933.82264 & -221.0 & 1.0 & HIRES \\
13933.82336 & -217.8 & 1.0 & HIRES \\
14085.16632 & -203.3 & 1.3 & HIRES \\
14085.16702 & -204.2 & 1.3 & HIRES \\
14085.16775 & -203.6 & 1.2 & HIRES \\
14130.12242 & 321.0 & 1.1 & HIRES \\
14130.12308 & 318.4 & 1.1 & HIRES \\
14130.12372 & 319.2 & 1.1 & HIRES \\
14131.11653 & 290.6 & 1.2 & HIRES \\
14131.11724 & 291.0 & 1.2 & HIRES \\
14131.11797 & 291.7 & 1.1 & HIRES \\
14139.10185 & 90.7 & 1.2 & HIRES \\
14139.10250 & 91.1 & 1.2 & HIRES \\
14139.10312 & 87.7 & 1.2 & HIRES \\
14641.84867 & -237.2 & 1.2 & HIRES \\
14807.16184 & 12.3 & 1.6 & HIRES \\
14809.16606 & 46.7 & 1.3 & HIRES \\
14811.16959 & 95.0 & 1.2 & HIRES \\
14865.09274 & -199.3 & 1.2 & HIRES \\
14866.06558 & -209.0 & 1.3 & HIRES \\
14868.10796 & -213.6 & 1.5 & HIRES \\                        
14927.00429 & 68.3 & 1.4 & HIRES \\
14963.76892 & -54.2 & 1.4 & HIRES \\
14983.92452 & -212.9 & 1.4 & HIRES \\
14984.93831 & -211.1 & 1.2 & HIRES \\
14985.93592 & -217.3 & 1.2 & HIRES \\
14986.92704 & -216.2 & 1.2 & HIRES \\
14987.92941 & -221.9 & 1.2 & HIRES \\
14988.87370 & -227.3 & 1.4 & HIRES \\
15016.74253 & -202.7 & 1.3 & HIRES \\
15042.73366 & 58.4 & 1.2 & HIRES \\
15043.74347 & 74.7 & 1.3 & HIRES \\
15044.77665 & 100.2 & 1.2 & HIRES \\
15256.94773 & -182.2 & 1.3 & HIRES \\
15286.01509 & 301.1 & 1.4 & HIRES \\
15311.80566 & -30.5 & 1.3 & HIRES \\
15312.79832 & -39.2 & 1.1 & HIRES \\
15314.85494 & -74.7 & 1.2 & HIRES \\
15375.75214 & -167.0 & 1.2 & HIRES \\
15405.73625 & 377.3 & 1.1 & HIRES \\
15585.18842 & -241.5 & 1.2 & HIRES \\
15607.18024 & -180.8 & 1.3 & HIRES \\                        
15636.97057 & 333.8 & 1.4 & HIRES \\
15636.97113 & 330.3 & 1.3 & HIRES \\
15636.97172 & 332.3 & 1.2 & HIRES \\
15636.97232 & 329.7 & 1.2 & HIRES \\
15706.73340 & -238.4 & 1.0 & HIRES \\
15707.73062 & -237.0 & 1.2 & HIRES \\
15731.94048 & -113.1 & 1.3 & HIRES \\
15792.72022 & -177.0 & 1.3 & HIRES \\
15792.72080 & -170.0 & 1.3 & HIRES \\
15792.72139 & -172.8 & 1.2 & HIRES \\
15993.02461 & 379.9 & 1.3 & HIRES \\
15993.02698 & 382.7 & 1.4 & HIRES \\
15993.03026 & 386.1 & 1.5 & HIRES \\
16145.72624 & -185.9 & 1.2 & HIRES \\
16488.73474 & -142.7 & 1.1 & HIRES \\
16675.18710 & 24.3 & 1.4 & HIRES
  \enddata
\end{deluxetable}


\subsection{Keplerian Orbital Solution}

\begin{figure*}
  \begin{center}
    \begin{tabular}{cc}
      \includegraphics[width=8.2cm]{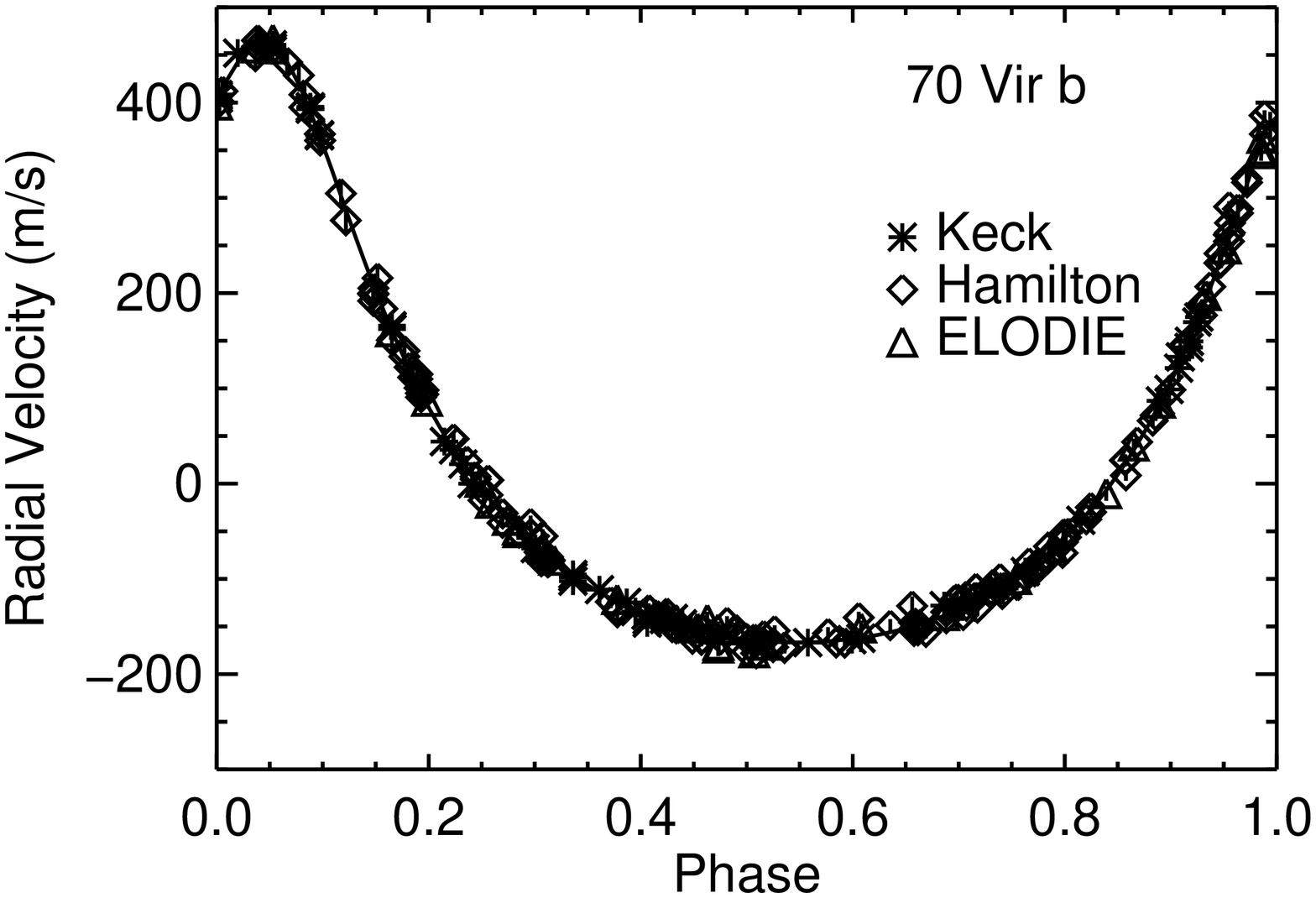} &
      \includegraphics[width=8.2cm]{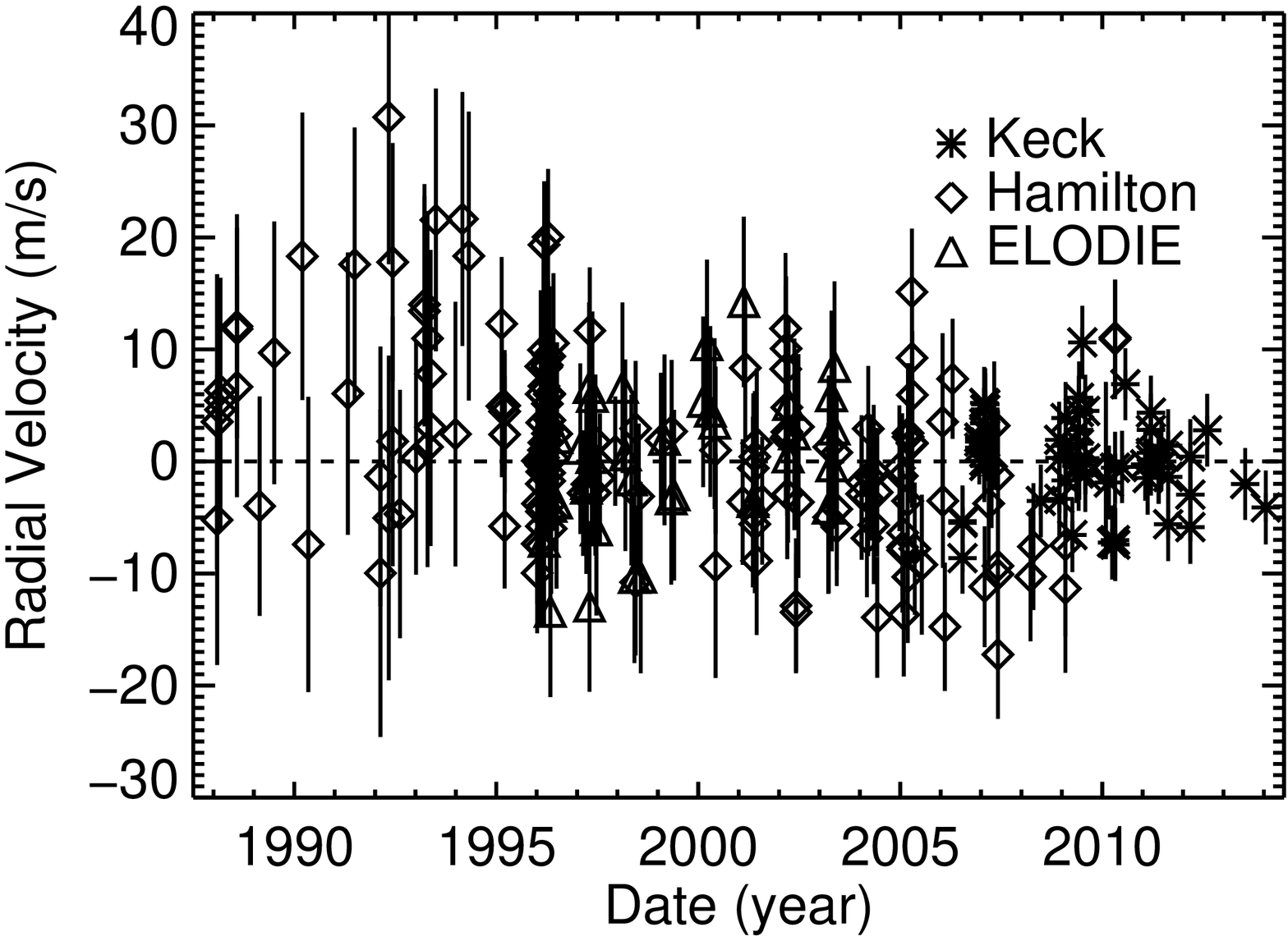}
    \end{tabular}
  \end{center}
  \caption{{\it Left}: All 263 RV measurements from three different
    instruments (see Table \ref{rvs}) for 70~Vir phased on the new
    orbital solution shown in Table \ref{planet}. RV offsets between
    datasets have been accounted for in this figure. {\it Right}:
    Residual velocities with respect to the best orbital solution.}
  \label{rv}
\end{figure*}

The revised Keplerian orbital solution to the RV data in Table
\ref{rvs} used RVLIN; a partially linearized, least-squares fitting
procedure described in \citet{wri09}. Parameter uncertainties were
estimated using the BOOTTRAN bootstrapping routines described in
\citet{wan12}. The resulting orbital solution is shown in Table
\ref{planet} and in Figure \ref{rv}.

\begin{deluxetable}{lc}
  \tablecaption{\label{planet} Keplerian Orbital Model}
  \tablewidth{0pt}
  \tablehead{
    \colhead{Parameter} &
    \colhead{Value}
  }
  \startdata
\noalign{\vskip -3pt}
\sidehead{70 Vir b}
~~~~$P$ (days)                    & $116.6926 \pm 0.0014$ \\
~~~~$T_c\,^{a}$ (JD -- 2,440,000) & $16940.258 \pm 0.084$ \\
~~~~$T_p\,^{b}$ (JD -- 2,440,000) & $7239.7091 \pm 0.11$ \\
~~~~$e$                           & $0.399 \pm 0.002$ \\
~~~~$\omega$ (deg)                & $358.8 \pm 0.3$ \\
~~~~$K$ (m\,s$^{-1}$)             & $315.7 \pm 0.7$ \\
~~~~$M_p$\,sin\,$i$ ($M_J$)       & $7.40 \pm 0.02$ \\
~~~~$a$ (AU)                      & $0.481 \pm 0.003$ \\
\sidehead{System Properties}
~~~~$\gamma$ (m\,s$^{-1}$)           & $22.94 \pm 0.59$ \\
\sidehead{Measurements and Model}
~~~~$N_{\mathrm{obs}}$            & 263 \\
~~~~rms (m\,s$^{-1}$)             & 6.08 \\
~~~~$\chi^2_{\mathrm{red}}$       & 1.16
  \enddata
  \tablenotetext{a}{Time of mid-transit.}
  \tablenotetext{b}{Time of periastron passage.}
\end{deluxetable}

For each bootstrapping realization, the fit produces offsets for each
dataset with respect to the Lick Hamilton dataset. These are fit as
two additional free parameters in the Keplerian orbital fit described
above. We find the offsets to be 48.4 and -74.6 m\,s$^{-1}$ for data
from ELODIE and HIRES respectively. The $\chi^2_{\mathrm{red}}$ and
rms scatter of the residuals (see Table \ref{planet}) are consistent
with the measurement uncertainties shown in Table \ref{rvs}. Note that
we added a stellar jitter noise component of 3~m\,s$^{-1}$ in
quadrature with the measurement uncertainties \citep{but06}. We find
no evidence for a linear RV trend in the fit residuals shown in the
right panel of Figure \ref{rv}.

\begin{figure*}
  \begin{center}
    \begin{tabular}{cc}
      \includegraphics[width=8.2cm]{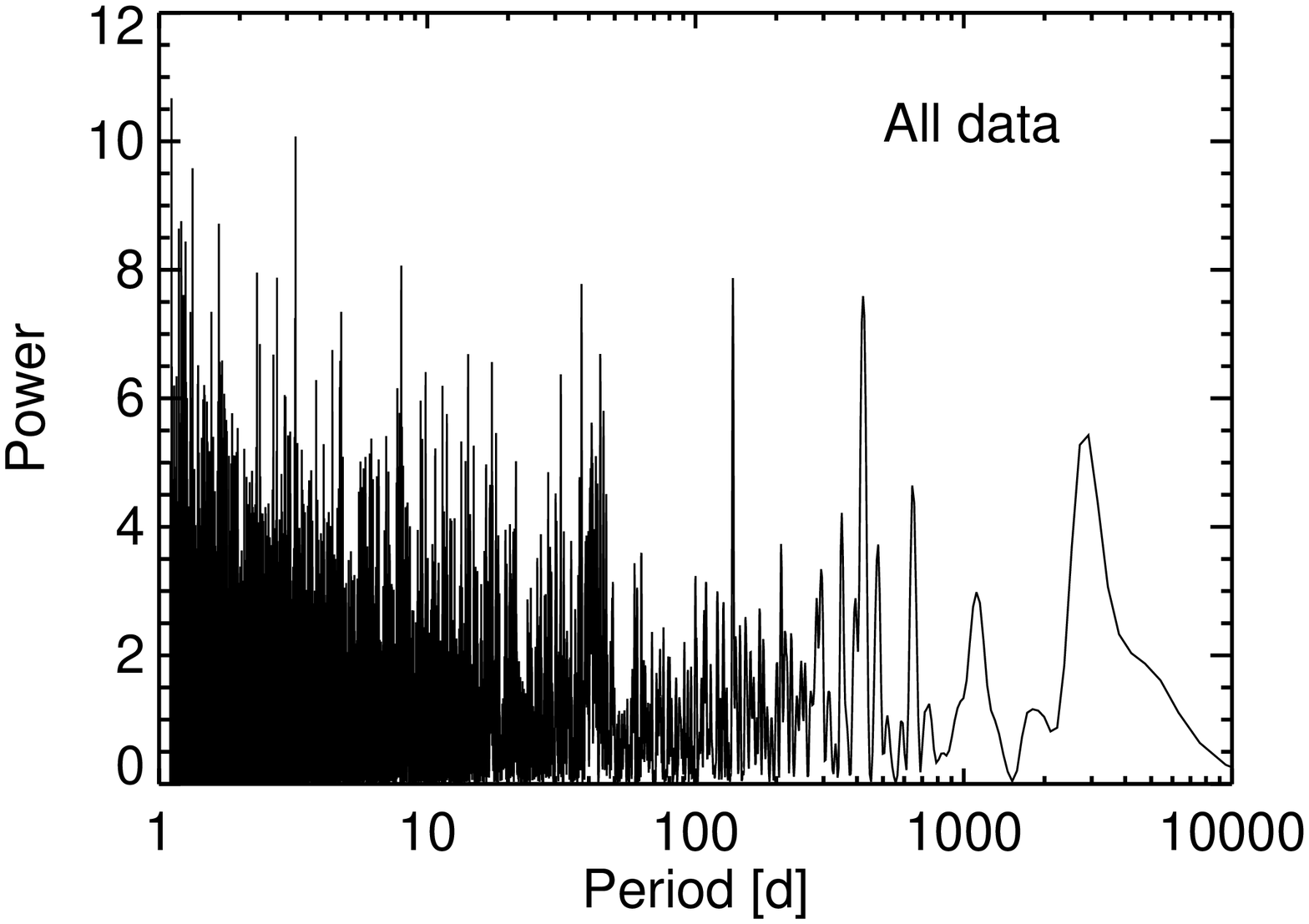} &
      \includegraphics[width=8.2cm]{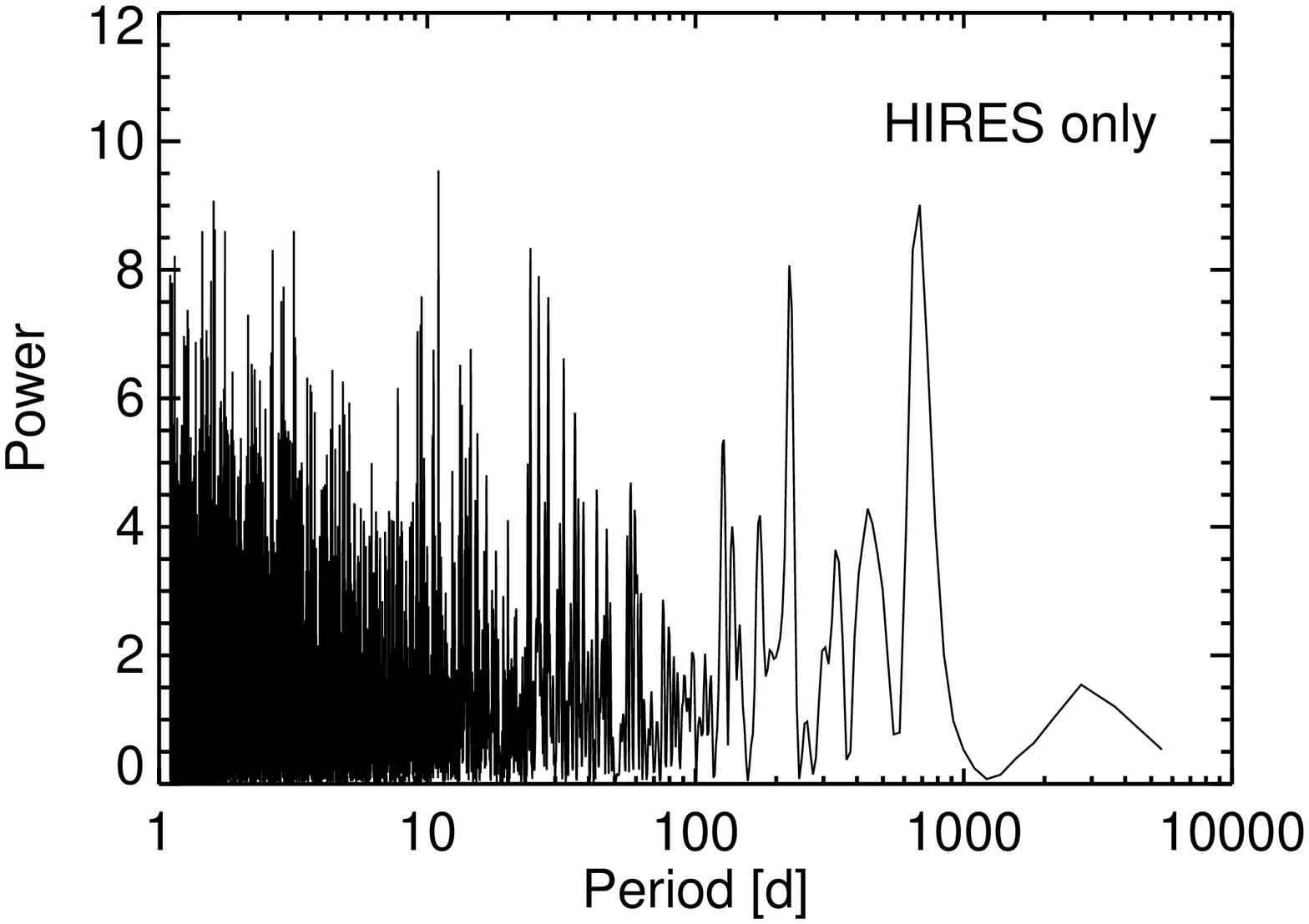}
    \end{tabular}
  \end{center}
  \caption{{\it Left}: A Lomb-Scargle periodogram of the residuals to
    the Keplerian orbital fit shown in Table~\ref{planet}, using all
    of the available data from Table~\ref{rvs}. The fourier powers
    shown in the figure are consistent with noise in the data.  {\it
      Right}: The same analysis repeated using only the HIRES data,
    showing that our data do not reveal the presence of additional
    planets.}
  \label{lsp}
\end{figure*}

We performed a further analysis of the data to search for signatures
of possible additional planets. Figure \ref{lsp} shows the
Lomb-Scargle periodogram \citep{hor86,sca82} of the best-fit
residuals, which shows no dominant peak. Since the lower-precision
ELODIE and Hamilton data might obscure a low-amplitude signal
detectable in the HIRES data, we have also examined the HIRES data
alone. Figure \ref{lsp} shows the periodogram of the residuals of the
HIRES data to the best fit shown in Table \ref{planet}.

The presence of many peaks of similar amplitude in these periodograms
is consistent with there being many, low-mass planets of similar RV
semi-amplitude in the data, but also consistent with noise. Since the
rms of the residuals is consistent with both expectations and
measurements of the uncertainties, there is no reason to expect the
former, so we conclude that these data contain no evidence of
additional periodic astrophysical signals.

\subsection{Transit Ephemeris Refinement}
\label{ephem}

The transit mid-point epoch shown in Table \ref{planet} was calculated
with a Monte-Carlo bootstrap, which propagates the uncertainty in this
orbital parameter to the time of the transit. This method produces the
most accurate ephemeris since the transit times are calculated as part
of the orbital fit. Note that if the planet does not transit then the
transit mid-point epoch may be considered the time of inferior
conjunction. The predicted transit properties of a system depend
sensitively on the stellar radius as well as the planetary
parameters. We adopt our interferometric measured radius from
Table~\ref{tab:stellar_properties} which has an uncertainty of only
1.5\%. The minimum mass of the planet is larger than a Jupiter mass
and we approximate the radius of the planet as $R_p = 1.0 R_J$, based
upon the mass-radius relationship described by \citet{kan12a}. These
properties, combined with the orbital solution of Table \ref{planet},
result in a transit probability of 2.27\%, a predicted transit
duration of 0.66~days, and a transit depth of 0.3\%. The transit
mid-point uncertainty shown in Table \ref{planet} is 0.084~days, or
121~minutes. Therefore the transit window is dominated by the transit
duration rather than the mid-point uncertainty, which is a favorable
scenario for photometric follow-up. Our procedure is to use a
calculated value for $T_c$ as close as possible to the conclusion of
observations. However, the baseline of the RV observations described
in Section \ref{sec:spectra} is long enough such that there is very
little increase in the size of this transit window for the foreseeable
future. The uncertainty in the predicted transit time subsequent to
that shown in Table~\ref{planet} has an uncertainty that is less than
a minute larger. \citet{kan08} have also shown that the transit
probability is a strong function of both the eccentricity and the
argument of periastron. For example, if the eccentricity of the planet
were zero, the transit probability and duration would be 1.92\% and
0.71~days respectively. Thus the orientation of the 70~Vir~b orbit
results in a slightly enhanced transit probability relative to a
circular orbit.


\section{System Habitable Zone}
\label{hz}

The fundamental stellar parameters from Table
\ref{tab:stellar_properties} provide the means to investigate the HZ
of the system and the potential for terrestrial planets in that
region. Previous studies of the 70~Vir HZ include those of
\citet{jon06}, who calculated HZ boundaries for a selection of known
exoplanetary systems using the estimated ages of stars to determine
on-going habitability. \citet{san07} studied stability regions in the
70~Vir system and concluded that the system is unlikely to host HZ
planets. These previous studies used the older HZ boundaries of
\citet{kas93}. Here we revisit the HZ properties of 70~Vir using the
revised system parameters presented here along with the updated HZ
calculations of \citet{kop13b,kop14}.

\begin{figure}
  \includegraphics[angle=270,width=8.5cm]{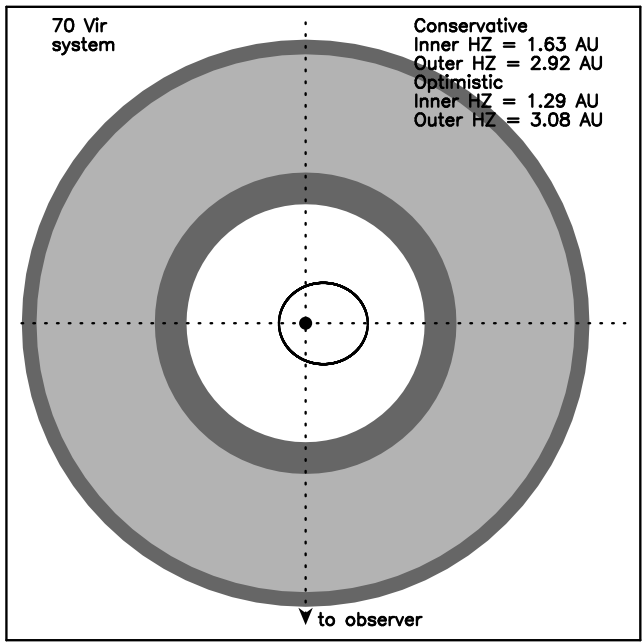}
  \caption{A top-down view of the 70~Vir system showing the extent of
    the HZ calculated using the stellar parameters of Table
    \ref{tab:stellar_properties}. The conservative HZ is shown as
    light-gray and optimistic extension to the HZ is shown as
    dark-gray. The revised Keplerian orbit of the planet from Table
    \ref{planet} is overlaid.}
  \label{hzfig}
\end{figure}

We adopt the definitions of ``conservative'' and ``optimistic'' HZ
models described by \citet{kan13}. The conservative HZ use boundaries
based upon runaway and maximum greenhouse climate models, whereas the
optimistic HZ extends these boundaries based on assumptions regarding
the amount of time that Venus and Mars were able to retain liquid
water on their surfaces \citep{kop13b}. The accuracy to which these
boundaries can be determined rely on robust determinations of the
stellar parameters \citep{kan14a} which, in this case, have
exceptionally small related uncertainties (see Table
\ref{tab:stellar_properties}) such that the HZ boundary uncertainties
are negligible. HZ calculations for all known exoplanetary systems are
available using the same methodology through the Habitable Zone
Gallery \citep{kan12a}.

Figure \ref{hzfig} shows a top-down view of the 70~Vir system where
the solid line indicates the Keplerian orbit of the planet using the
orbital parameters of Table \ref{planet}. The HZ is depicted by the
shaded region where the light gray represents the conservative HZ and
the dark gray is the optimistic extension to the HZ. The conservative
HZ covers the region 1.63--2.92~AU from the host star and the
optimistic HZ increases this region to 1.29--3.08~AU.

\begin{figure*}
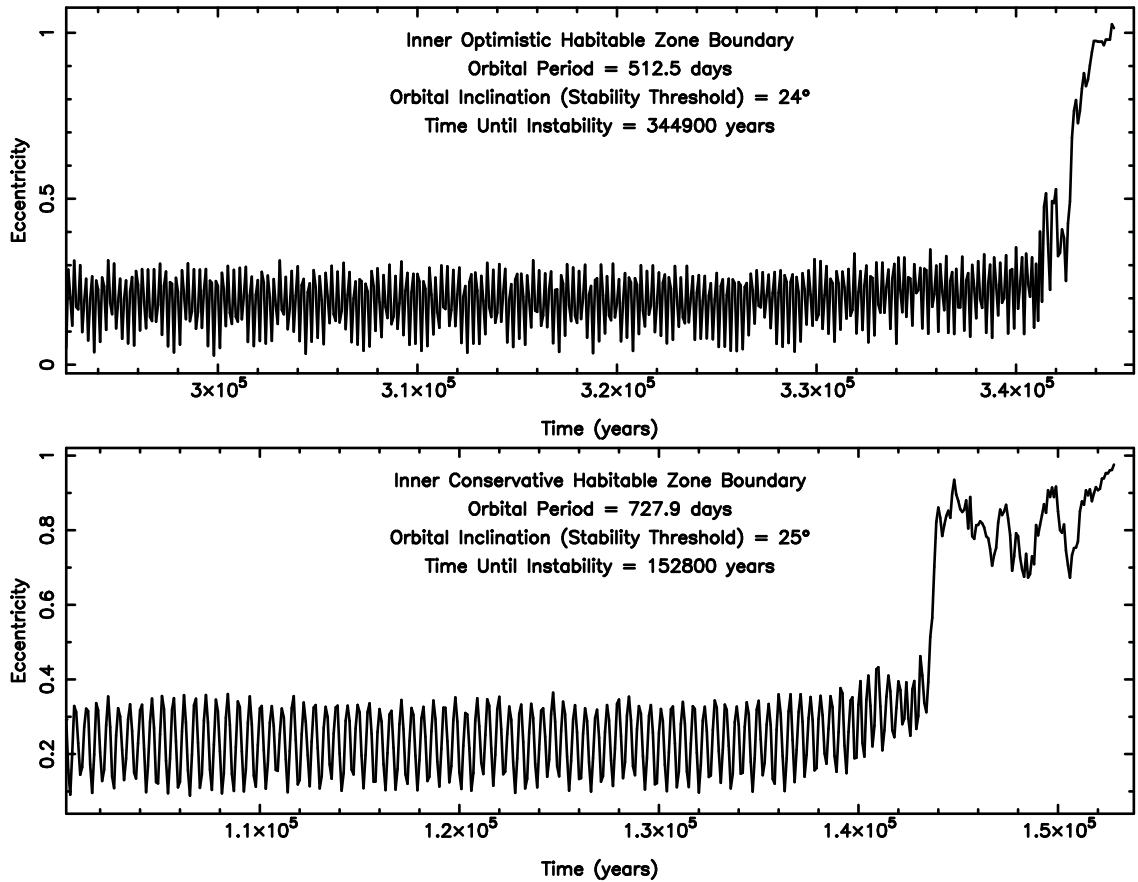

  \begin{center}
    \includegraphics[angle=270,width=15.0cm]{f06a.ps} \\
    \includegraphics[angle=270,width=15.0cm]{f06b.ps}
  \end{center}
  \caption{Stability simulations for a hypothetical Earth-mass planet
    in the HZ of the 70~Vir system. Each panel shows the eccentricity
    oscillations for the 50,000 years leading up to the ejection of
    the outer planet. Top panel: The Earth-mass planet remains stable
    at the inner boundary of the optimistic HZ for system inclinations
    $> 24\degr$, otherwise the inner planet causes the outer planet to
    be ejected after $\sim$~350,000 years. Bottom panel: The
    Earth-mass planet remains stable at the inner boundary of the
    conservative HZ for system inclinations $> 25\degr$, otherwise the
    inner planet causes the outer planet to be ejected after
    $\sim$~150,000 years.}
  \label{hzstabfig}
\end{figure*}

Although the confirmed planet is clearly interior to the HZ, we
performed stability simulations to investigate whether the relatively
large mass of the planet and the eccentricity of its orbit exclude the
presence of a hypothetical Earth-mass planet in the HZ. To accomplish
this, we performed dynamical simulations using N-body integrations
with the Mercury Integrator Package \citep{cha99}. We adopted the
hybrid symplectic/Bulirsch-Stoer integrator and used a Jacobi
coordinate system, which provides more accurate results for
multi-planet systems \citep{wis91,wis06} except in cases of close
encounters \citep{cha99}. We inserted the hypothetical planet in a
circular orbit at each of the four optimistic and conservative HZ
boundaries. The integrations were performed for a simulation of $10^6$
years, in steps of 100 years, starting at the present epoch.

Assuming that the system is coplanar with an inclination of $90\degr$,
our simulations show that the hypothetical systems all remain stable
for the full duration of the simulations. The eccentricity of the
hypothetical planet oscillates over the course of the simulation with
a range of 0.00--0.35, 0.04--0.30, 0.05--0.22, and 0.03--0.18 for the
optimistic inner, conservative inner, conservative outer, and
optimistic outer boundaries respectively. These eccentricities do not
necessarily rule out habitability of the planet, depending on the
dynamics of the planetary atmosphere \citep{kan12b,wil02}. Note that
stability at the HZ boundaries does not guarantee stability within the
boundaries as that is a complex function of orbital distance, phase,
and eccentricity.

Since the mass of the inner planet depends on the inclination of the
system ($M_p \sin i = 7.40 M_J$), we performed simulations that
determine the system inclination where the mass of the inner planet
causes the orbit of the outer planet to become unstable. These
stability threshold inclinations for the four boundaries are
$24\degr$, $25\degr$, $10\degr$, and $3\degr$ for the optimistic
inner, conservative inner, conservative outer, and optimistic outer
boundaries respectively. Shown in Figure \ref{hzstabfig} are the
simulation results at the stability threshold inclination for the
inner optimistic and conservative HZ boundaries. Each panel shows the
eccentricity oscillations for the 50,000 years leading up to the
instability event. Even though the inner conservative HZ boundary is
farther away from the inner planet, the orbital period at that
boundary places the outer planet closer to an orbital resonance with
the inner planet than an orbit at the inner optimistic HZ
boundary. Thus the planet remains stable for less time at the former
than the latter.


\section{Photometric Observations}
\label{photometry}

We have been monitoring 70~Vir for two decades with the T4 0.75~m
automatic photoelectric telescope (APT) located at Fairborn
Observatory in southern Arizona. The T4 APT observes in the
Str\"omgren $b$ and $y$ pass bands with an EMI 9124QB photomultiplier
tube (PMT) as the detector. The automated photometer has a Fabry lens
placed behind the focal-plane diaphragm that projects a fixed image of
the primary mirror (illuminated by the star) onto the photo-cathode of
the PMT. Thus, slight motions of the star within the diaphragm during
an integration do not translate into image motion on the PMT
cathode. The instrumentation and observing strategy result in the data
being close to the photon/scintillation noise limit with far less
correlated noise than is typical of CCD photometry which suffers from,
for example, intra-pixel sensitivity. The data are thus assumed to be
uncorrelated in the subsequent analysis. As an additional verification
of validity of this assumption, each of the APT integrations are
divided into a series of 0.1 second integrations and saved for quality
control and trouble shooting purposes. Histograms of the subinterval
data and computed Geneva statistics are used to verify that the data
are Gaussian and determine if there are trends, cycles, spikes, or
drops in the photon counts during the integration that require further
investigation. The T4 APT, its photometer, observing techniques, data
reduction procedures, and photometric precision are described in
further detail in \citet{h1999}.

The comparison star HD~117304 (C1: $V=5.65$, $B-V=1.05$, K0~III) has
been used for all 23 observing seasons since 1993, while comparison
star HD~112503 (C2: $V=6.81$, $B-V=0.47$, F7~IV) has been used only
for the past 19 observing seasons beginning in 1997 because it was
chosen to replace a previous comparison star recognized to be a
low-amplitude variable after the first four years. T4 has acquired
2051 good differential observations with C1 over the past 23 years and
1897 good observations with C2 in the past 19 observing
seasons. During the course of our analysis, we recognized that
comparison star C1 exhibited very low-amplitude variability at times;
therefore, in this paper, we present the results of our analysis of
the 1897 differential magnitudes of 70~Vir with respect to comparison
star HD~112503 (C2).

\begin{figure}
  \includegraphics[width=8.2cm]{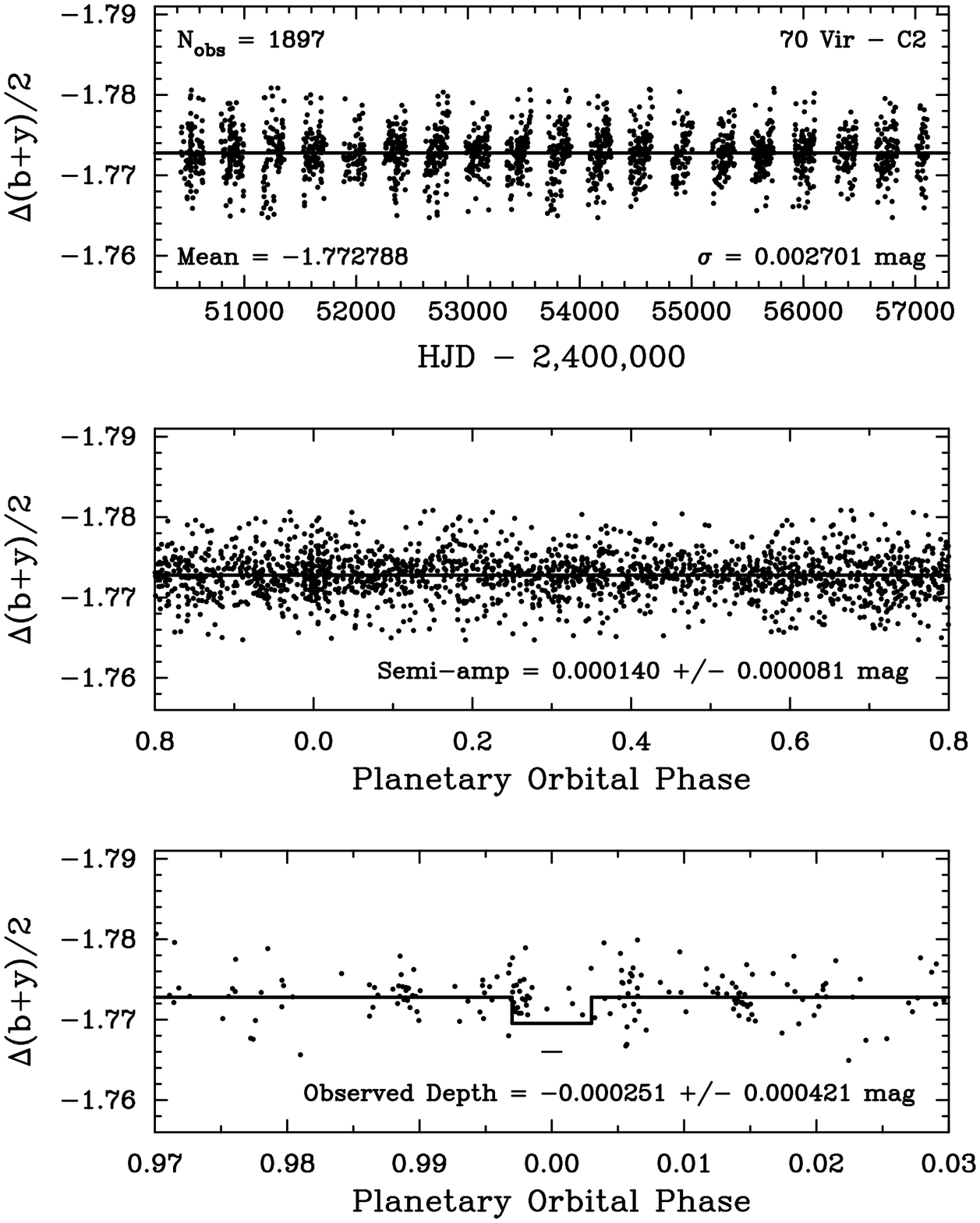}
  \caption{{\it Top}: The 1997--2015 normalized differential
    observations of 70~Vir with respect to comparison star C2,
    acquired with the T4 0.75~m APT between 1997 and 2015. {\it
      Middle}: The observations phased with the 116.6926 day orbital
    period. The semi-amplitude of a least-squares sine fit to the
    phase observations is $0.000140\pm0.000081$~mag, consistent with
    the absense of light variability on the radial velocity period and
    confirming planetary reflex motion of the star as the cause of the
    radial velocity variations. {\it Bottom}: The observations within
    $\pm0.03$ phase units of the predicted transit time. The solid
    curve shows the predicted transit time at phase 0.0, transit depth
    (0.3\%) and duration ($\pm0.003$ phase units) for a central
    transit of planet b. The short horizontal line segment represents
    the uncertainty in the time of tansit. Our photometry shows that
    central transits of the expected depth and duration likely do not
    occur.}
  \label{photfig}
\end{figure}

The 1897 differential magnitudes computed with C2 are plotted in the
top panel of Figure \ref{photfig}. To increase the precision of the
differential magnitudes, we combined the $b$ and $y$ observations into
a single $(b+y)/2$ "pass band." We also normalized each observing
season to have the same mean magnitude as the first, thus making our
search for short-period variability and shallow transits more
sensitive. The nightly normalized observations scatter about their
grand mean, indicated by the straight line in the top panel, with a
standard deviation 0.00270~mag. This is slightly larger than our
typical measurement precision given above and may indicate slight
residual variability in 70~Vir and/or the comp star HD~112503.

The observations are replotted in the middle panel of Figure
\ref{photfig}, where they have been phased with the time of
conjunction and the orbital period from Table \ref{planet}. A
least-squares sine fit on the 116.6926-day radial velocity period
gives a formal semi-amplitude of just $0.000352\pm0.000076$ mag, thus
confirming that the periodic radial velocity variations are due to
planetary reflex motion and not to intrinsic stellar brightness
variations \citep[see, e.g.,][]{qhs+2001,psc+2004,bbs2012}.

The observations within $\pm0.03$ phase units of the predicted transit
time are plotted in the bottom panel of Figure \ref{photfig}. The
solid curve shows the predicted transit phase (0.0), the transit depth
(0.3\% or 0.00325~mag), and transit duration ($\pm0.003$ phase units)
computed as described in Section \ref{ephem} above. The horizontal
line below the transit window represents the transit mid-point
uncertainty. While the second half of the transit window is not well
covered by our observations, there are a total of 27 observations
within the transit window that have a mean of
$-1.773035\pm0.000416$~mag and 1870 observations that fall outside the
transit window and have a mean of $-1.772784\pm0.000063$~mag. Thus,
our "observed" transit depth is $-0.000251\pm0.000421$~mag, which is
consistent with zero to three decimal places. Therefore, a central
transit of the expected depth and duration occurring at the expected
time can be ruled out at the 5$\sigma$ level. Although the data
sampling is sufficient to also constrain the absence of transits for
almost all impact parameters, the number of data points within the
corresponding transit durations will be less, thereby lessening the
significance of such constraints. For example, the largest gap in the
photometry during the transit window corresponds to $\sim$0.3 of the
central transit duration. This means the impact parameter would need
to be $\geq 0.95$ to have been completely missed by our data. Ruling
out such a range of impact parameters would reduce the posterior
transit probability from 2.27\% to 0.11\%.

70~Vir is a magnetically inactive star with $\log{R'_{HK}}$ values of 
-4.99 and -5.116 according to \citet{wmb+2004} and \citet{if2010}, 
respectively. \citet{wmb+2004} give an estimated rotation period of 32 days 
for 70~Vir, based on the star's activity level. However, no reliable 
rotation period for 70 Vir has been directly measured via rotational 
modulation of dark starspots or bright Ca~II~H~and~K regions across 
the face of the star \citep[e.g.,][]{hen97,hbd+2000,sim10}. We 
performed periodogram analyses of each individual observing season and of 
our data set as a whole, and, while we found suggestions of low-amplitude
variability, we could not identify any significant period that might be 
interpreted as the stellar rotation period. 

\begin{figure}
  \includegraphics[width=8.2cm]{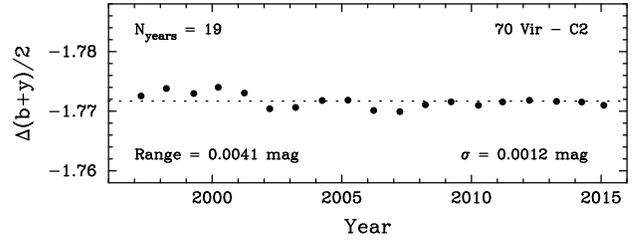}
  \caption{The 1997--2015 yearly mean differential magnitudes 70~Vir
    minus comparison star C2. The error bars on the individual
    seasonal means are slightly smaller than the plotted points. The
    19 differential magnitudes vary over a total range of 0.0041~mag
    and have a standard deviation of 0.0012~mag with respect to the
    grand mean, indicated by the horizonal dotted line. Without
    another good comparison star, we are unable to determine whether
    the variability is intrinsic to 70~Vir or the comparison star, so
    these values can only be quoted as upper limits of long-term
    variability in 70~Vir.}
  \label{photmean}
\end{figure}

Finally, we look for long-term variability in 70~Vir. Unfortunately,
comparison star C1 has significant long-term variability of several
mmag. The yearly-mean differential magnitudes (70~Vir - C1) vary over
a range 0.0076 mag and have a standard deviation of 0.0023 mag with
respect to the grand mean. However, the yearly means of (70~Vir - C2)
have both a smaller range and a smaller standard deviation, 0.0041~mag
and 0.0012~mag, respectively (see Figure \ref{photmean}). Without
another good comparison star, we cannot determine whether the
variability we see in Figure \ref{photmean} originates in 70~Vir, the
comparison star, or a combination of both. Thus, we can only state
that 70~Vir's long-term variability has a range less than $\sim$0.004
mag.


\section{Conclusions}
\label{conclusions}

In an era where new planets are being regularly discovered via the
transit method, the bright exoplanet host stars still largely belong
to those planets which were discovered using the radial velocity
method. These are systems which thus provide the greatest access to
follow-up investigations due to the relatively large signal-to-noise
possibilities presented. Here we have presented new results for the
70~Vir system which includes detailed characterization of the host
star. Our direct measurements of the stellar radius show that,
although slightly cooler, 70~Vir is almost twice the size of the
Sun. This is consistent with the star being older and more evolved
than the Sun. Our new radial velocity data provide an improved
Keplerian orbital solution for the planet and further evidence that
there are unlikely to be further giants planets within the system. A
terrestrial-mass planet may yet exist beneath our detection threshold
and so, given the vastly improved stellar properties, we calculated
the HZ boundaries and performed stability simulations within that
region. Our simulations show that a terrestrial planet could remain in
a stable orbit near the HZ inner edge for system inclinations $>
25\degr$ and close to the outer HZ edge for almost all system
inclinations. Finally, our 19 years of APT photometry confirm that the
star is quite stable over long time periods and there is no evidence
that the b planet transits the host star (a ``dispositive null'', as
described by \citet{wan12}), the timing of which we were able to
accurately predict from our revised Keplerian orbit. The TERMS
compilation of data for this system presented here means that it is
now one of the better characterized systems in terms of stellar and
planetary parameters.


\section*{Acknowledgements}

We thank the anonymous referee for helpful comments which improved the
manuscript. S.R.K and N.R.H. acknowledge financial support from the
National Science Foundation through grant AST-1109662. T.S.B
acknowledges support provided through NASA grant ADAP12-0172. The
CHARA Array is funded by the National Science Foundation through NSF
grants AST-0606958 and AST-0908253 and by Georgia State University
through the College of Arts and Sciences, as well as the W. M. Keck
Foundation. G.W.H. acknowledges support from NASA, NSF, Tennessee
State University, and the State of Tennessee through its Centers of
Excellence program. Y.K.F. and J.T.W. acknowledge support from NASA
Keck PI Data Awards, administered by the NASA Exoplanet Science
Institute, including awards 2007B N095Hr, 2010A N147Hr, 2011A\&B
N141Hr, \& 2012A N129Hr; NASA Origins of Solar Systems grant
NNX09AB35G; NASA Astrobiology Institute grant NNA09DA76A; and the
Center for Exoplanets and Habitable Worlds (which is supported by the
Pennsylvania State University, the Eberly College of Science, and the
Pennsylvania Space Grant Consortium). This research has made use of
the Habitable Zone Gallery at hzgallery.org. This research has also
made use of the SIMBAD and VIZIER Astronomical Databases, operated at
CDS, Strasbourg, France (http://cdsweb.u-strasbg.fr/), and of NASA's
Astrophysics Data System, of the Jean-Marie Mariotti Center
\texttt{SearchCal} service (http://www.jmmc.fr/searchcal),
co-developed by FIZEAU and LAOG/IPAG.


\end{document}